\documentclass{article}
\usepackage{spconf,amsmath,amssymb,graphicx,hyperref}
\usepackage{booktabs,multirow,tabularx,makecell}
\usepackage{url}
\usepackage[T1]{fontenc}
\usepackage[utf8]{inputenc}
\usepackage{microtype}
\usepackage{inconsolata}
\usepackage{xcolor,orcidlink}
\usepackage{comment}

\title{ESDCodec: Ultra-Low Bitrate Speech Codec with Enhanced Semantic Disentanglement and Subband Decoding}
\name{
Nan Xu~\orcidlink{0009-0006-1010-0381},
Kecan Mao~\orcidlink{0009-0007-7936-3603},
Xin Gu~\orcidlink{0000-0003-4375-1456},
Yi Zhao~\orcidlink{0000-0002-3555-9408}$^{*}$\thanks{$^{*}$Corresponding author. Email:athenazhao@tencent.com},
Zhi Gong~\orcidlink{0009-0002-1850-3848}}
\address{WeChat, Tencent Inc, Beijing, China}

\begin{document}
\ninept
\maketitle

\begin{abstract}
\sloppy
Despite recent advances in neural speech codecs, achieving high-fidelity speech reconstruction at low bitrates remains a formidable challenge. To address this limitation, we propose ESDCodec, a speech codec that integrates a thoroughly enhanced semantic quantizer and a conditioned decoder network. Specifically, we employ a randomly initialized and frozen codebook, followed by a lightweight projector, to encode semantic details entirely within a linear space while enhancing codebook utilization. To further improve perceptual quality, we design a condition network that explicitly injects prior subband knowledge into the upsampling decoder. Taking the de-quantized feature as input, this network predicts subband signals, thereby providing fine-grained guidance for waveform reconstruction. Extensive experimental results show that ESDCodec achieves superior reconstruction performance at an ultra-low bitrate of 0.85kbps. For LLM-based speech generation tasks, ESDCodec also consistently outperforms existing codec models.

\end{abstract}


\section{Introduction}

\sloppy
Neural speech codecs serve as a foundational technology for speech signal compression and have garnered increasing research interest in recent years. The primary objective of a speech codec is to compress speech signals into compact discrete representations while preserving perceptual fidelity. With the rapid progress of large language models (LLMs), the integration of speech modalities into text-based LLMs has become more prevalent. In such scenarios, neural speech codecs play a pivotal role in bridging continuous speech waveforms with discrete tokens. Leveraging these discrete representations, speech language models (SLMs) have been successfully applied to various downstream tasks, including speech understanding, text-to-speech (TTS) synthesis, and spoken dialogue systems. 

\sloppy
In practice, there exists an inherent trade-off between bitrate and reconstruction quality. Recent advances in quantization techniques, such as residual vector quantization (RVQ) \cite{juang1982multiple}, have enabled flexible bitrate configurations. However, when the bitrate drops to approximately 1 kbps, reconstruction performance often deteriorates significantly \cite{defossez2022high}. To mitigate this issue, single-codebook codecs such as Single-Codec \cite{li2024single} and WavTokenizer \cite{ji2024wavtokenizer} have been proposed, compressing all information into a single vector quantization (VQ) sequence. Additionally, SimVQ \cite{zhu2025addressing} introduces a frozen codebook paradigm with global updates through a linear layer, achieving full use of the codebook. Despite these innovations, such methods often lack explicit semantic modeling, limiting their effectiveness in semantics-aware applications.

\sloppy
Semantic tokens are typically derived from self-supervised learning (SSL) models or supervised models \cite{chen2022wavlm}. For example, SpeechTokenizer \cite{zhang2023speechtokenizer}  distills semantic information into the first RVQ layer using an SSL model, while the X-Codec series \cite{ye2025codec} explicitly injects SSL semantic features into the acoustic stream. Acoustic tokens, on the other hand, are commonly obtained from codecs optimized for acoustic detail, such as SoundStream \cite{zeghidour2021soundstream}, EnCodec \cite{defossez2022high}, and Descript-Audio-Codec (DAC) \cite{kumar2023high}. Although these models achieve high fidelity across various bitrates, they often suffer from misaligned semantic content, resulting in suboptimal compatibility with SLMs. Recent advancements in dual-stream codec architectures have explored the disentanglement of semantic and acoustic tokens. The Mimi codec \cite{defossez2024moshi} distills semantic features into a single VQ module. SemantiCodec \cite{liu2024semanticodec} employs AudioMAE \cite{huang2022masked} to achieve the disentanglement of semantic and acoustic streams. DualCodec \cite{li2025dualcodec} thoroughly decouples semantic and acoustic quantizers into the unified dual-stream architecture. Although these codecs achieve exceptional reconstruction quality, low bitrates still hinder further improvements. Integrating semantic-acoustic modeling with low-bitrate objectives into a cohesive framework remains an open challenge.

\begin{figure*}[t]
\begin{minipage}[b]{1.0\linewidth}
  \centering
  \centerline{\includegraphics[width=13.5cm]{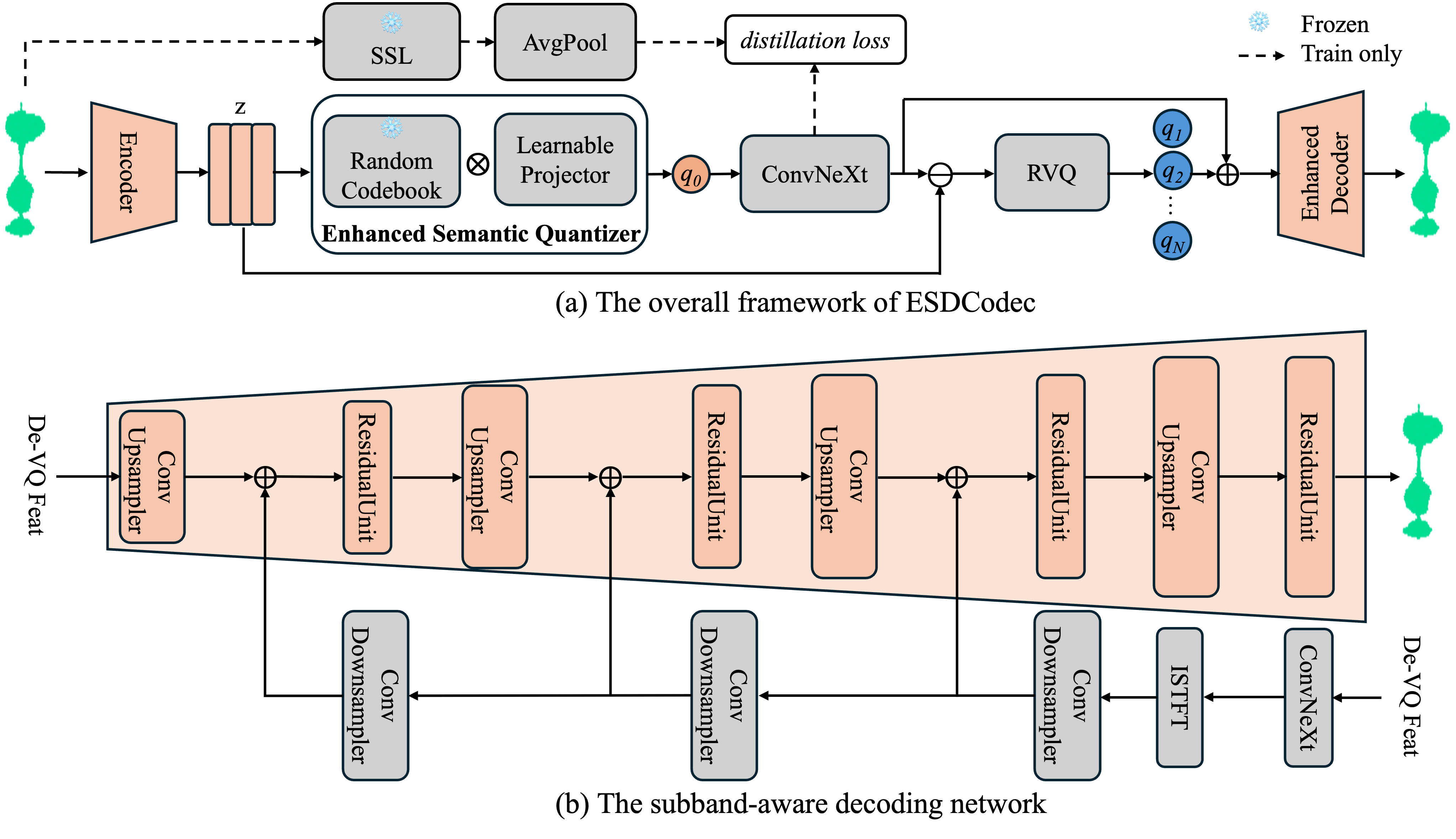}}
\end{minipage}
\caption{The overall architecture of the proposed ESDCodec. In the subfigure (a), the enhanced semantic quantizer is shown and generates the single semantic code $ q_0 $. The remaining acoustic codes are encoded by RVQ layers. The dashed line refers to the SSL distillation loss and corresponding modules only involve during the training process. In the subfigure (b), the detailed framework of the subband-aware decoding network is shown. The de-quantized features are used as inputs for upsampling and condition networks, respectively.}
\label{frmw}
\end{figure*}

\sloppy
To overcome these limitations, we propose ESDCodec, a dual-stream neural speech codec featuring an enhanced semantic quantizer and a conditioned decoder. The semantic quantizer adopts a frozen codebook with a learnable linear projector, inspired by SimVQ, to promote full codebook activation and mitigate representation collapse. This module is trained with SSL distillation and operates with architectural simplicity during inference. Furthermore, we enhance the decoder with a subband condition network that injects prior spectral knowledge into each upsampling stage. By explicitly modeling subband structures, the enhanced decoder achieves more accurate waveform reconstruction. The main contributions of this work are summarized as follows:
\sloppy
\begin{itemize}
\item We propose ESDCodec, a dual-stream framework that effectively disentangles semantic and acoustic tokens for low-bitrate neural speech codecs. The enhanced semantic quantizer, combining a frozen codebook and learnable projector, ensures efficient semantic encoding with minimal bitrate overhead.
\item We introduce a subband-aware decoder network that leverages prior subband information to guide waveform reconstruction. This design significantly improves perceptual quality and reconstruction accuracy.
\item Extensive experiments demonstrate that ESDCodec achieves state-of-the-art performance at 0.85kbps, outperforming existing baselines in both objective and subjective evaluations. We also release our code and pre-trained model to support reproducibility and future research\footnote{\url{https://github.com/vspeech/ESDCodec}}.
\end{itemize}

\section{Method}

\subsection{Overall framework}
\sloppy
The overall architecture of ESDCodec is illustrated in Figure~\ref{frmw}(a), which adopts a dual-stream quantization paradigm with distinct semantic and acoustic pathways. In the encoding stage, the raw waveform is mapped to a latent space $ Z\in \mathbb{R} ^{T\times D} $ via a convolutional encoder. This encoder performs a 960-fold temporal compression, reducing the original 24 kHz sampling rate to a 25 Hz frame-level representation. Built upon stacked downsampling convolutional layers with snake activations \cite{ziyin2020neural}, the encoder outputs features that are then routed to an enhanced semantic quantizer. This quantizer yields merely a single token $ q_0 $, thereby substantially reducing the overall bitrate. To enhance the discriminative power of the semantic branch, we introduce an SSL distillation objective, drawing on the feature representations from both w2v-BERT-2.0 \cite{barrault2023seamless} and ConvNeXt \cite{liu2022convnet}. Concurrently, the supplementary acoustic information is progressively encoded via residual vector quantization layers, consistent with the design principles of DAC \cite{kumar2023high}.
 

\subsection{Enhanced semantic quantizer}
\sloppy
The codebook collapse is a typical issue and plagues the traditional vector quantization \cite{zhu2025addressing}. To efficiently quantize the semantic information, we introduce the SimVQ \cite{zhu2025addressing} technique to enhance the single-layer semantic quantizer. By leveraging full codebook utilization and global space updates, the enhanced semantic quantizer generates compact and informative semantic codes. Specifically, the codebook is reparameterized as a combination of a randomly initialized and frozen basis matrix $ S\in \mathbb{R} ^{N\times F} $ and a learnable linear projection matrix $ W\in \mathbb{R} ^{F\times D} $, which can be written as follows,
\begin{equation}
\label{eq1}
\setlength{\abovedisplayskip}{6pt}
\setlength{\belowdisplayskip}{6pt}
C= S\times W.
\end{equation}

\sloppy
Instead of learning individual code vectors, the entire codebook $ C\in \mathbb{R} ^{N\times D} $ is jointly optimized during training and only $ W $ is updated. This design prevents representation collapse and encourages full codebook utilization. We can obtain the quantized embedding through nearest-neighbor lookup in the codebook $ C $. Specifically for each frame $ z_t $, we have
\begin{equation}
\label{eq2}
\setlength{\abovedisplayskip}{6pt}
\setlength{\belowdisplayskip}{6pt}
i_{t} =arg\min_{n} \left || z_{t} -c_{n}  \right || _{2}^{2},
\end{equation}
where $ c_{n}\in C $. To capture richer semantic structures, we place ConvNeXt \cite{liu2022convnet} modules immediately after the quantization operation, drawing on the design of \cite{li2025dualcodec}. Simultaneously, we introduce an extra L2 self-supervised distillation loss to preserve the informational integrity of the semantic embeddings throughout training.

\begin{table*}[t]
\renewcommand{\arraystretch}{1.2}
\centering
\caption{Reconstruction results around 0.9kbps bitrate on LibriSpeech test-clean subset. The codebook size with $ x/y $ denotes semantic and acoustic streams in the dual-stream codec, respectively. Nq refers to the number of used quantizers. Bold indicates the best scores.}
\label{tab1}
\resizebox{1.0\textwidth}{!}{
\setlength{\tabcolsep}{1.2mm}{
\begin{tabular}{*{2}{l}*{11}{c}}
  \toprule
  {{\bf Codec Model}} & {{\bf Params}} & {{\bf Bit(kbps)}} & {{\bf Tok/s}} & {{\bf Codebook size}} & {{\bf Nq}} & {{\bf PESQ$\uparrow$}} & {{\bf STOI$\uparrow$}} & {{\bf V/UV F1$\uparrow$}} &{{\bf Pitch$\downarrow$}} & {{\bf ViSQOL$\uparrow$}} &{{\bf WER$\downarrow$}} & {{\bf MUSHRA$\uparrow$}} \\  
  \midrule
  {Ground-Truth}& - & - & - & - & - & 4.64 & 1.000 & - & -  & - & 3.70 & 91.6 \\  
   \midrule
  {DAC 75Hz}& 74.71M & 0.75 & 75 & 1024 & 1 & 1.17 & 0.737 & 0.744 & 271.6  & 1.50 & 35.3 & 28.7 \\ 
  {Encodec 75Hz}& 14.85M & 1.50 & 150 & 1024 & 2 & 1.56 & 0.845 & 0.836 & 134.7  & 1.94 & 6.68 & 37.4  \\ 
  {SpeechTokenizer 50Hz} & 103.68M & 1.00 & 100 & 1024 & 2 & 1.25 & 0.777 & 0.802 & 179.1 & 1.34  & 5.35 & 35.1 \\ 
  {X-Codec 50Hz}& 79.50M & 1.00 & 100 & 1024 & 2 & 2.11  & 0.863 & 0.913 & 52.39 & 2.36 & {\bf 3.75} & 83.4 \\ 
  {DualCodec 25Hz}& 100.0M & 0.85 & 75 & 16384/1024 & 3 & 2.32  & 0.912 & 0.929 & 38.76  & 2.61 & 4.70 & 85.8 \\ 
  {ESDCodec 25Hz}& 89.78M & 0.85 & 75 & 16384/1024 & 3 & {\bf 2.49}  & {\bf 0.921} & {\bf 0.936} & {\bf 33.86} & {\bf 2.74} & 4.18 & {\bf 88.2} \\  
  \bottomrule
\end{tabular}}}
\end{table*}

\subsection{Subband-aware decoding}
\sloppy
The decoder constitutes a critical module in neural audio codecs, responsible for the conversion of compressed embeddings into audible signals of high perceptual quality. In standard implementations, this conversion is realized through a stack of upsampling convolutional layers. However, the direct mapping from low-dimensional quantized features to time-domain waveforms is inherently ill-posed, as previously underscored in the vocoder literature \cite{xu2025universal,siuzdak2023vocos,xu2026}.

\sloppy
To overcome this limitation, we propose the enhanced codec decoder equipped with a subband-aware condition network. As illustrated in Figure~\ref{frmw}(b), the de-quantized continuous features are passed through four convolutional upsampling blocks and ResidualUnit \cite{zeghidour2021soundstream} modules to reconstruct the waveform. Simultaneously, the condition network employs ConvNeXt blocks to predict the magnitude and phase of subband signals. The inverse STFT is then applied to generate subband waveforms. Furthermore, this subband prior information is integrated into the codec decoder via subsequent downsampling layers, where outputs from each upsampling block are combined with corresponding downsampling features. The subband-aware condition network is lightweight, adding only 9.6M parameters.

\subsection{Training objective}
\sloppy
The training objective comprises adversarial, reconstruction, SSL distillation, and quantization losses.

\noindent{\bf Adversarial loss.} We adopt the adversarial loss $ \mathcal{L}_{adv} $ from DAC and employ the L1 feature matching loss $ \mathcal{L}_{feat} $ for intermediate features between ground truth and generated samples. For discriminators, we use the multi-scale STFT and multi-period discriminators that are consistent with DAC.

\noindent{\bf Reconstruction loss.} The reconstruction loss $ \mathcal{L}_{recon} $ contains three losses. We use the multi-scale Mel spectrogram losses $ \mathcal{L}_{full} $ and $ \mathcal{L}_{sub} $ for reconstructed fullband and subband audios, respectively. Additionally for the condition network, we use a single STFT loss \cite{yang2021multi} $ \mathcal{L}_{stft} $ to constrain magnitudes between the predicted and ground truth samples.

\noindent{\bf SSL distillation loss.} Semantic distillation is performed via an L2 loss $ \mathcal{L}_{distill} $ between features from a pre-trained SSL model and those reconstructed by ConvNeXt.

\noindent{\bf Quantization loss.} The quantization loss $ \mathcal{L}_{vq} $ contains two losses, i.e., codebook loss and commitment loss. The codebook loss $ \mathcal{L}_{code} $ is employed between quantized embeddings and encoder outputs. The commitment loss $ \mathcal{L}_{commit} $ is also used to remain encoder outputs close to assigned codebook entries. These losses are used for both semantic and acoustic streams.

The overall training objective can be written as follows,
\begin{multline}
\label{eq3}
\setlength{\abovedisplayskip}{6pt}
\setlength{\belowdisplayskip}{6pt}
\mathcal{L}=\lambda _{adv} \mathcal{L}_{adv}+\lambda _{feat} \mathcal{L}_{feat}\\
+\lambda _{recon} \mathcal{L}_{recon}+\lambda _{distill} \mathcal{L}_{distill}+\lambda _{vq} \mathcal{L}_{vq},
\end{multline}
where hyperparameters balance each loss component.

\section{Experiments}
\subsection{Datasets and training setups}
\sloppy
\noindent{\bf Datasets.} To ensure diversity in the training data, we adopt approximately 97,000 hours of bilingual (English and Chinese) speech from the Emilia dataset \cite{he2024emilia}, all resampled to 24 kHz. To evaluate the generalization capability of the proposed codec, we employ the out-of-domain LibriSpeech \verb+test-clean+ set \cite{panayotov2015librispeech}, consisting of 2,620 utterances, as our test corpus. 

\sloppy
\noindent{\bf Implementation.} In ESDCodec, we use the encoder module in DAC \cite{kumar2023high}, containing 4 convolution blocks with strides (4, 5, 6, 8). Therefore for 24kHz waveform input, we can generate 25Hz frame-rate tokens. The decoder without the condition network mirrors the encoder and is composed of stacked temporal upsampling convolution layers. For acoustic RVQ module, we use 11 codebook layers. The codebook size is 1024 and codebook dim is 8. For semantic quantization module, the frozen codebook size is set as 16384 and codebook dim is set as 1024. The input and output channels of the lightweight projector layer are both set as 1024. Additionally, the ConvNeXt network following the semantic quantization module contains 4 blocks with the input dim 512 and the intermediate dim 1536. In the enhanced codec decoder, the condition network utilizes 6 ConvNeXt blocks to predict the waveform of a quarter subband. The FFT size, window size and hop size are 960, 960 and 240, respectively. The input dim and intermediate dim in each ConvNeXt block are set as 256 and 1024, respectively. Additionally, the strides of three downsampling convolution layers are set as 1, 5 and 4.

\sloppy
\noindent{\bf Training.} During training, we randomly intercept 4 seconds for each speech sample and employ the batch size as 12. Notably, we utilize the weight normalization for convolution modules. The initial learning rates for both the generator and discriminator are set as 1e-4 and the exponential decay is employed throughout training. AdamW optimizer with betas (0.8, 0.9) is employed for model optimization. Finally for all experiments, we train all codec models for 350K steps and TTS models for 600K steps. Notably, we use 8 codebooks for TTS training and all models are trained on 8 NVIDIA H20 GPUs.

\subsection{Baselines and evaluations}

\sloppy
\noindent{\bf Baselines.} Our baselines are composed of a series of advanced codec models, including RVQ-based codecs (DAC \cite{kumar2023high} and Encodec \cite{defossez2022high}), semantic enhanced codecs (SpeechTokenizer \cite{zhang2023speechtokenizer} and X-Codec \cite{ye2025codec}), and dual-stream codec DualCodec \cite{li2025dualcodec}. Notably, in our experiments, we utilize public official implementations and checkpoints.

\sloppy
\noindent{\bf Evaluations.} In our experiments, both objective and subjective evaluations are used as metrics for our proposed codec and baselines. We employ six different evaluations as objective metrics, including the Perceptual Evaluation of Speech Quality (PESQ) \cite{rix2001perceptual} with wide band\footnote{\url{https://github.com/ludlows/python-pesq}}, Short-Time Objective Intelligibility (STOI)\footnote{\url{https://github.com/mpariente/pystoi}}, Word Error Rate (WER), two pitch-related metrics and ViSQOL\footnote{\url{https://github.com/google/visqol}} \cite{chinen2020visqol}. Notably, for the WER metric, transcriptions are obtained with \verb+Paraformer-zh+ for Chinese and \verb+Whisper-large-v3+ for English. F1 score of voiced/unvoiced classification (V/UV F1) and pitch error using F0 Root Mean Square Error\footnote{\url{https://github.com/gemelo-ai/vocos}} are used as pitch-related metrics. Additionally, we use MUSHRA \cite{schoeffler2018webmushra} listening tests for the subjective evaluation. Specifically, we recruit 10 participants to complete the test using randomly selected 15 utterances from LibriSpeech dataset.

\begin{table}[t]
\renewcommand{\arraystretch}{1.2}
\centering
\caption{The TTS experiment results of different codec models on Seed-TTS-Eval benchmark. Evaluation metrics of models* are reported by DualCodec. Bold denotes the best scores.}
\label{tab2}
\resizebox{0.46\textwidth}{!}{
\setlength{\tabcolsep}{4.3mm}{
\begin{tabular}{*{1}{l}*{4}{c}}
  \toprule
  \multirow{2}*{{\bf Codec Model}} & \multicolumn{2}{c}{{\bf WER$\downarrow$}} & \multicolumn{2}{c}{{\bf SIM$\uparrow$}} \\  
  \cmidrule(lr){2-3}\cmidrule(lr){4-5}
  &  {{\bf EN}} & {{\bf ZH}} & {{\bf EN}} & {{\bf ZH}} \\
  \midrule
  {Ground-Truth} & 2.13 & 1.25 & 0.73 & 0.75  \\
  {SpeechTokenizer*}& 15.4 & 21.5 & 0.47 & 0.55 \\
  {DualCodec*}& 3.40 & 2.49 & 0.57 & 0.67  \\
  {ESDCodec}& {\bf 3.31} & {\bf 2.39} & {\bf 0.63} & {\bf 0.70}  \\
  \bottomrule
\end{tabular}}}
\end{table}

\subsection{Results}
\sloppy
\subsubsection{Codec performance}
As illustrated in Table~\ref{tab1}, we compare all codecs under a similar ultra-low bitrate (around 0.85kbps) on LibriSpeech \verb+test-clean+ set. Overall, the proposed ESDCodec delivers the best reconstruction quality while maintaining a comparable bitrate and token rate compared to other low-bitrate systems. Specifically, ESDCodec achieves the highest objective metrics among all baseline codecs, obtaining a PESQ of 2.49 and ViSQOL of 2.74. It also yields superior pitch-related performance with the best F1 score and the lowest pitch error, indicating more accurate voicing decisions and pitch tracking at ultra-low bitrates. Compared to the closely matched dual-stream baseline DualCodec that has the same bitrate/token rate and codebook partition, ESDCodec improves PESQ by 0.17, STOI by 0.01, and ViSQOL by 0.13, while reducing pitch error by 4.9, demonstrating that the proposed method better preserves both intelligibility and quality given the same token budget. In terms of semantic content reconstruction, although X-Codec achieves the lowest WER, it does so at a higher bitrate. Under the stricter low-bitrate setting (0.85kbps), ESDCodec achieves a competitive WER of 4.18\%. Furthermore, ESDCodec achieves the best subjective listening results among all baseline codecs, which aligns well with the improvements observed in objective measures and suggests that the proposed design yields perceptually meaningful gains at ultra-low bitrates. 

\subsubsection{TTS performance}
\sloppy
We train the VALL‑E model \cite{wang2023neural} on discrete codec tokens generated by various codecs, and evaluate its TTS performance on the Seed‑TTS‑Eval benchmark\footnote{\url{https://github.com/BytedanceSpeech/seed-tts-eval}}. The word error rate (WER) and speaker similarity (measured by SIM‑O, abbreviated as SIM) are reported in Table~\ref{tab2}. Overall, ESDCodec achieves superior TTS quality over all competing codec‑based tokenizers, consistently improving both content accuracy and speaker similarity for English and Chinese.

\subsubsection{Ablation study}
\sloppy
Table~\ref{tab3} presents an ablation study to assess the contribution of key design choices in ESDCodec. The full model achieves the best overall performance across all metrics, indicating that the complete architecture yields the most favorable trade-off between perceptual quality, intelligibility, and prosody fidelity. Furthermore, replacing the enhanced semantic pathway with the VQ-based semantic stream (i.e., w/ VQ semantic) consistently degrades reconstruction and linguistic accuracy, while also slightly harming voicing and pitch accuracy. This suggests that the semantic full-utilization representation in ESDCodec provides more informative and better-aligned conditioning for the decoder than a vector quantized semantic alternative under the same operating setting. Additionally, removing the conditioning mechanism (i.e., w/o condition) causes the largest performance deterioration, particularly for semantic content and prosody. These results indicate that the conditioning is crucial for preserving content-related cues and stabilizing prosodic reconstruction at the ultra-low bitrate, thereby improving both perceptual quality and alignment with the semantic content.

\begin{table}[t]
\renewcommand{\arraystretch}{1.2}
\centering
\caption{The ablation experiment results on LibriSpeech test-clean subset. The best results are listed in bold.}
\label{tab3}
\resizebox{0.46\textwidth}{!}{
\setlength{\tabcolsep}{1.3mm}{
\begin{tabular}{*{1}{l}*{5}{c}}
  \toprule
  {{\bf Method}} & {{\bf PESQ$\uparrow$}} & {{\bf STOI$\uparrow$}} & {{\bf V/UV F1$\uparrow$}} &{{\bf Pitch$\downarrow$}} & {{\bf WER$\downarrow$}}  \\  
  \midrule
  {ESDCodec}& {\bf 2.49} & {\bf 0.921} & {\bf 0.936} & {\bf 33.86} & {\bf 4.18}  \\
  {w/ VQ semantic}& 2.43 & 0.919 & 0.932 & 34.02 & 4.43  \\ 
  {w/o condition}& 2.42 & 0.916 & 0.930 & 36.21 & 6.38  \\ 
  \bottomrule
\end{tabular}}}
\end{table}

\begin{figure}[t]
\includegraphics[width=\columnwidth]{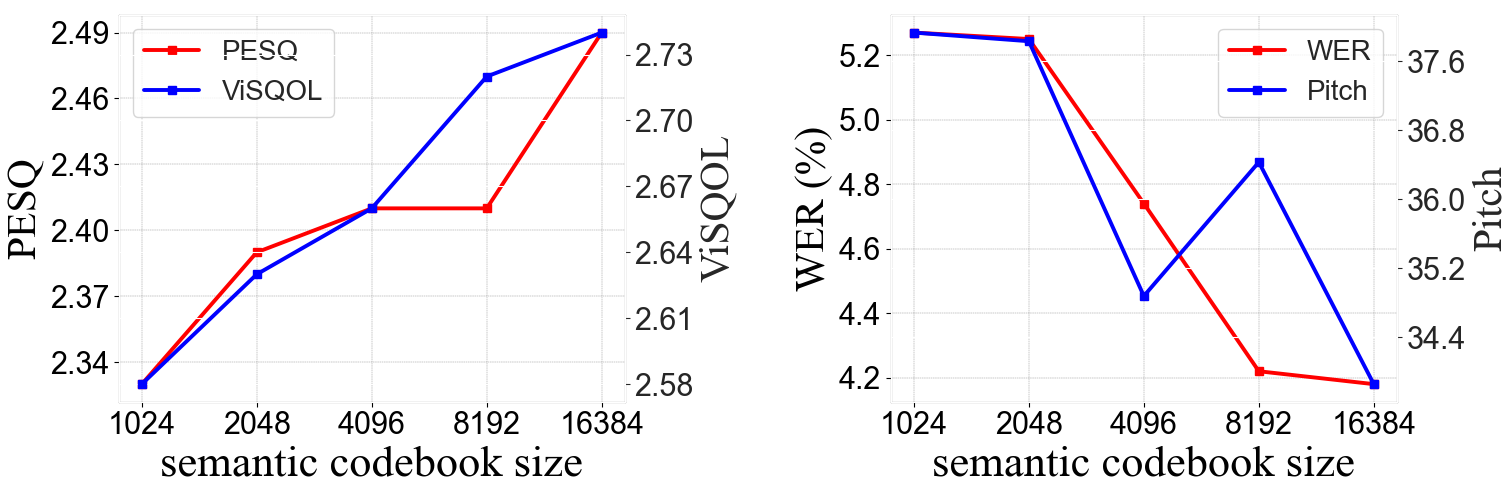}
\caption{The ablation experiment results of different semantic codebook sizes on LibriSpeech test-clean subset.}
\label{a1}
\end{figure}

\sloppy
As illustrated in Figure~\ref{a1}, we present an ablation study on the impact of the size of the semantic codebook on ESDCodec. As the codebook size increases from 1024 to 16384, we observe consistent improvements in both perceptual quality and content preservation. In particular, ViSQOL increases from 2.58 to 2.74, indicating better reconstruction fidelity with larger semantic codebooks. Meanwhile, the word error rate (WER) decreases from 5.27\% to 4.18\%, demonstrating enhanced semantic retention. These gains can be attributed to the expanded representation capacity of a larger codebook, which allows finer-grained modeling of semantic information and reduces quantization errors.

\section{Conclusion}
\sloppy
In this paper, we introduce ESDCodec, a dual-stream neural speech codec designed for high-fidelity waveform reconstruction at ultra-low bitrates. The proposed enhanced semantic quantizer, which combines a frozen codebook with a learnable linear projector, mitigates the codebook collapse and ensures efficient semantic encoding with minimal bitrate overhead. Additionally, the subband-aware decoder, equipped with a lightweight condition network, injects prior spectral knowledge into the reconstruction process, significantly improving the perceptual quality. Extensive experiments on LibriSpeech demonstrate that ESDCodec achieves state-of-the-art performance at 0.85kbps, outperforming existing codecs across both objective and subjective metrics. Furthermore, ablation studies validate the contributions of each proposed component. 


\bibliographystyle{IEEEbib}
\bibliography{mybib}

\end{document}